# Functional Equivalence Checking for Verification of Algebraic Transformations on Array-Intensive Source Code

K.C. Shashidhar[1,2], Maurice Bruynooghe[2], Francky Catthoor[1,2] and Gerda Janssens[2]

[1]Interuniversitair Micro-Elektronica Centrum (IMEC) vzw, Kapeldreef 75, B-3001 Heverlee, Belgium
[2]Faculteit Toegepaste Wetenschappen, Katholieke Universiteit Leuven, Belgium
{kodambal,catthoor}@imec.be, {maurice,gerda}@cs.kuleuven.ac.be

**Abstract**

*Development of energy and performance-efficient embedded software is increasingly relying on application of complex transformations on the critical parts of the source code. Designers applying such nontrivial source code transformations are often faced with the problem of ensuring functional equivalence of the original and transformed programs. Currently they have to rely on incomplete and time-consuming simulation. Formal automatic verification of the transformed program against the original is instead desirable. This calls for equivalence checking tools similar to the ones available for comparing digital circuits. We present such a tool to compare array-intensive programs related through a combination of important global transformations like expression propagations, loop and algebraic transformations. When the transformed program fails to pass the equivalence check, the tool provides specific feedback on the possible locations of errors.*

## 1 Introduction

Source code transformations come into play in situations where a designer wants much better optimizations than those a compiler can provide. Such a situation is common for designers of mobile computing and communicating systems. They are required to program complex signal processing algorithms for complex platform architectures and yet meet stringent constraints on the energy consumption and performance of the final implementation. Research has shown that application of source-to-source code transformations on an original implementation of the algorithm can greatly help in meeting such constraints (cf. [3, 4, 5, 13]).

Every stage in an implementation activity brings forth an associated verification problem. Source code transformations are no exception. The problem here is to ensure that the transformed program preserves the functionality of the original program. Designers are at present using simulation of the transformed program to gain confidence in the correctness of the transformed program before forwarding it to the synthesis stage. But simulation is both incomplete and time-consuming. Also, when the transformed program is in error, it is hard to detect the exact cause with simulation. Clearly, formal automatic verification of the transformed program against the original, with support for error diagnostics, is desirable here. A pragmatic approach to this problem is to separate the two concerns, viz., applying transformations and verifying that they preserve the functional equivalence. This implies an *ex post facto* solution that requires a program equivalence checking tool. Our work addresses this requirement. Since, in general, the program equivalence problem is undecidable, we target the most important transformations applied on a decidable, and yet relevant, class of programs.

**Code transformations considered.** We are interested in verification of source code transformations that reduce the accesses to the data memory hierarchy. Broadly, there are two kinds of such transformations viz., *global loop transformations* and *global data-flow transformations*. Global loop transformations are applied to reorder and restructure the `for`-loops in the complete program in order to minimize the data transfers between different layers of the hierarchy by improving the temporal and spatial locality of the accessed data. On the other hand, global data-flow transformations are applied either to remove repeated computation or to break bottlenecks caused by data-flow dependencies in the program. They comprise of *expression propagations* that introduce or eliminate temporary variables that hold intermediate values and *global algebraic transformations* that take advantage of algebraic properties of the operators in transforming the data-flow. The need for verification support for these transformations is rather high because they invariably involve error prone manipulation of the index expressions of the array variables, especially when applied manually.

We do not distinguish between the transformations as long as they are only from the above categories. The transformed program can be under a combination of the transformations. The equivalence checking is done *oblivious* of any information about the particular instances of the above transformations that were applied and the order of their application.

**An example problem.** Suppose that we are given program functions, as in Fig. 1. Expression propagations and loop transformations have been applied on the original function (a) to obtain (b), and additionally, algebraic transformations to obtain (c) and (d). The functions, when executed, take inputs `A[]` and `B[]`, and assign the computed values to the elements of the output array variable `C[]` and terminate. If we ignore the possibility of overflow in the evaluation of fixed-point integer expressions, the integer addition is both associative and commutative. Therefore, it is expected that, if the same values are input to the functions, the same values are assigned to the elements of the output variable `C[]`.

```
/* Original function */
#define N 1024
foo(int A[], int B[], int C[])
{
    int k, tmp[N], buf[2*N];

    for(k=0; k<N; k++)
s1: tmp[k] = B[2*k] + B[k];

    for(k=N; k>=1; k--)
s2: buf[2*k-2] = A[2*k-2]
                 + A[k-1];
    for(k=0; k<N; k++)
s3: C[k] = tmp[k] + buf[2*k];
}
```
**a**

```
/* Transformed function ver 1 */
#define N 1024
foo(int A[], int B[], int C[])
{
    int k, tmp[N], buf[N];

    for(k=0; k<512; k++)
t1: tmp[k] = B[2*k] + B[k];

    for(k=0; k<N; k++){
t2:   buf[k] = A[2*k] + A[k];
      if (k < 512)
t3:      C[k] = tmp[k] + buf[k];
      else
t4:      C[k] = (B[2*k] + B[k])
                + buf[k];
    }
}
```
**b**

```
/* Transformed function ver 2 */
#define N 1024
foo(int A[], int B[], int C[])
{
    int k, buf[2*N];

    for(k=0; k<N; k++)
u1: buf[k] = A[k] + B[k];

    for(k=N; k<=2*N-2; k+=2)
u2: buf[k] = A[k] + B[k];

    for(k=0; k<N; k++)
u3: C[k] = buf[k] + buf[2*k];
}
```
**c**

```
/* Transformed function ver 3 */
#define N 1024
foo(int A[], int B[], int C[])
{
    int k, tmp[N], buf[2*N];

    for(k=0; k<=2*N-2; k+=2)
v1: buf[k] = A[k] + B[k];

    for(k=1; k<N; k+=2)
v2: tmp[k] = A[k] + B[k];

    for(k=0; k<N-1; k+=2){
v3: C[k] = buf[k] + buf[k];
v4: C[k+1] = tmp[k+1]
             + buf[2*k+2];
    }
}
```
**d**

**Fig. 1: Program functions (a), (b) and (c), equivalent under the considered transformations, compute** $\forall k \in [0\ldots N-1]$ : C[k] = B[2*k] + B[k] + A[2*k] + A[k]. **Program (d) is erroneously obtained. It is inequivalent to them** $\forall$ even $k \in [0\ldots N-1]$**, where** C[k] = A[k] + B[k] + A[k] + B[k]**, but equivalent** $\forall$ odd $k \in [0\ldots N-1]$.

That is, the functions are *input-output equivalent* under the applied transformations. We have developed a tool to check such equivalences fully automatically. Because of an erroneous transformation, (d) is not equivalent to (a), (b) and (c). It is helpful if the reason for nonequivalence of the function can be ascertained and debugged. To this end, our tool provides diagnostics when it fails to show an equivalence.

## 2 Related Work

Motivated by pragmatics, we are interested in a fully automatic, push-button style, a posteriori solution. This precludes discussion of the vast research on formal verification of the transformation tool or the library of transformations.

Undecidability of the program equivalence problem enforces that any effort start by defining a decidable class of programs that is of interest. Hence, the problem has been addressed by various researchers for different program classes with different applications in mind. Without enumerating the methods, to the best of our knowledge, none of the methods is able to show equivalence of program functions as in our example, in a scalable way. The problem we address is distinct by its central requirement to represent and maintain the relationships among elements of the array variables in the programs in closed form. Unrolling deeply nested loops with large bounds is clearly infeasible for real-life signal processing programs. To add to this, algebraic transformations will require a prohibitive search for normalization on the unrolled statements. Hence, we restrict our discussion of related work to methods that do not propose unrolling of loops.

Translation validation [7, 10] and fractal symbolic analysis (FSA) [9], both present methods which show semantic equivalence of two versions of programs. In the case of the former, the comparison is between the source and the target code. These methods are distinct from ours in that they essentially try to heuristically *infer* a sequence of legal transformations that can relate the two programs. Instead, we are able to directly check for equivalence of programs that are in a suitable language class. Also, their methods do not handle algebraic transformations. The work most related to ours, because we address the same class of programs, is the algorithm recognition method presented in [2]. Again, algebraic data-flow transformations are not handled by them. Another distinction is that, all these methods do not stress on debugging support which is very important in the context of source code transformations.

With respect to the equivalence problem of algebraic expressions, we would like to point out that methods from symbolic algebra do not suffice due to the presence of loop transformations on array variables. The quest here is for an analysis that deals with algebraic transformations involving array variables in the expressions.

We have presented a preliminary method that checks equivalence under only expression propagations and loop transformations in [11]. In this paper, based on an improved program representation, we present an extended method that is able to handle algebraic data-flow transformations, in combination with the other two transformations, in a single pass.

## 3 Program Representation

In this section we present a representation that captures the computation and the relationships between elements of array variables in the program. The representation is possible only when the programs belong to a class that we describe first.

### 3.1 Class of allowed programs

The class of programs that we consider is based on the recurring features in signal processing programs, and the availability of some tools (mentioned below) to convert programs that are not in the class. We assume that a program is first subject to a source-to-source code preprocessing phase in order to ensure that the original program has the properties of this class *before* beginning to apply the code transformations. Restriction of programs to this class is primarily to ease the analyses required to identify optimization opportunities and apply transformations. Apparently, what eases transformations also eases their verification.

The following properties distinguish the class: ① *Single-assignment form*: Programs have been converted to a form (called the *dynamic*-single assignment form) where every memory location is written only once (methods exist to automate this, for example, [6]); ② *Static control-flow*: Data-dependent `while`-loops have been converted to `for`-loops with worst-case bounds by moving the conditions inside the loops, and the data dependent `if`-conditions are simple enough to be handled by if-conversions; ③ *Affine indices*: All expressions in the index and the loop bounds are either affine or piece-wise affine, and ④ *No pointer references*: Pro-grams have been converted to a form where all references to the memory are with explicit indexing (for example, using a method as in [12]).

### 3.2 Array data dependence graphs

Given that a program has the above mentioned properties, we can extract the complete data-flow in the program. This extracted data-flow can be represented in the form of an *array data dependence graph* (ADDG). It is a directed graph, where, nodes represent the variables and *occurrences* of operators/functions in the program functions, and edges represent the *data dependence* (in the direction opposite to the *flow* of data). For example, Fig. 2 gives the ADDGs of the program functions in Fig. 1. The edges outgoing from operator nodes are labeled by the *position* of the operand for the computation and the thick edges outgoing from the variables are labeled by the labels of assignment statements where they are defined.

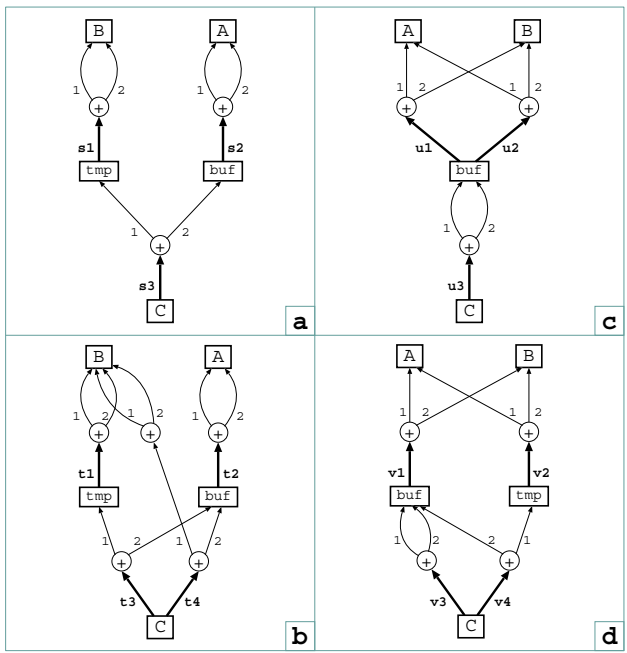

**Fig. 2: The ADDGs of the program functions in Fig. 1.**

The distinction when compared to a standard data dependence graph is that, in an ADDG, the data dependence, denoted by a reverse directed edge, refers not just to a single value, but to a *set of values*. Since the program is required to be in single-assignment form, the values are guaranteed to be assigned to different elements of the array variable defined in the statement. The dependence relation between the sets of values of defined variables and the operand variables is given by the so-called *dependency mappings*, one for each operand variable of the statement. For example, for the statement s2 in the original function (a), the two dependency mappings viz., from buf[] to the first A[] and from buf[] to the second A[] are as defined below, where $D := \{[k] \mid 1 \leq k \leq 1024 \wedge k \in \mathbb{Z}\}$.

$$M_{\text{buf},A_1} := \{[x] \to [y] \mid x = 2k - 2 \wedge y = 2k - 2 \wedge k \in D\}$$
$$M_{\text{buf},A_2} := \{[x] \to [y] \mid x = 2k - 2 \wedge y = k - 1 \wedge k \in D\}.$$

When there are multiple occurrences of a variable in the same statement we distinguish them with subscripts denoting their position as operands. In an ADDG, the root nodes and the leaf nodes correspond to the output and the input variables, respectively, of the program function. The rest of the variable nodes in the ADDG are intermediate variables.

**Intermediate variable reduction.** For a given path, reduction of an intermediate variable is an operation that we use as a primitive in our method. This involves updating the dependency mapping from the predecessor variable node to the intermediate variable node being reduced. The new dependency mapping will then be from the predecessor variable node to the successor variable node on the path in question. This is obtained by the composition of the two mappings.

For example, let us consider reduction of the intermediate variable node tmp on the leftmost path (path 1) in the ADDG of the original function (a) from the output variable C to the input variable B, that is, $[\ \text{C} \xrightarrow{s3} + \xrightarrow{1} \text{tmp} \xrightarrow{s1} + \xrightarrow{1} \text{B}\ ]$. The predecessor variable node to tmp is C and the successor variable node is B. Reducing tmp on the path involves updating the dependency mapping from C to tmp ($M_{\text{C,tmp}}$) into dependency mapping from C to B ($M_{\text{C} \overset{1}{\leadsto} \text{B}}$). It is computed as below, where $\bowtie$ is the natural join operation on two relations [8] and $D := \{[k] \mid 0 \leq k < 1024 \wedge k \in \mathbb{Z}\}$.

$$\begin{aligned} M_{\text{C} \overset{1}{\leadsto} \text{B}} &:= M_{\text{C,tmp}} \bowtie M_{\text{tmp},B_1} \\ &:= \{[k] \to [k] \mid k \in D\} \bowtie \{[k] \to [2k] \mid k \in D\} \\ &:= \{[k] \to [2k] \mid k \in D\}. \end{aligned}$$

On a given path, if all the intermediate variables between the current variable and the output variable are reduced, we obtain the *output-current mapping*. When the current variable is an input variable, this mapping gives the relation between the elements of the output variable to the elements of the input variable for that path. Then it is called the *output-input mapping* for that path of computation.

In the example path we mentioned above, tmp is the only intermediate variable, hence the output-input mapping from C to B for that path is just the $M_{\text{C} \overset{1}{\leadsto} \text{B}}$ that was computed above.

## 4 Algebraic Transformations

Algebraic data-flow transformations take advantage of the properties of the operators or user-defined functions and modify the data-flow such that the semantics of the original function are preserved (modulo overflow). The algebraic transformations are not restricted to the expression in a statement, but can have a global scope. This can be seen in our simple example in Fig. 1, where the algebraic transformations applied on (a) to obtain (c) and (d) are across expressions of multiple assignment statements at an algorithmic level. The ADDGs of the functions, as shown in Fig. 2, also reflect this.

Typically, most of such global transformations just rely on the associativity and/or commutativity properties of the operators like addition and multiplication on a fixed-point datatype like integer. Hence in what follows, we restrict our discussion to only these transformations. Other algebraic properties related to identity, inverse, distributivity and evaluation of constants are less common in practice and can be handled in a way similar to what we present.

The effect of such algebraic transformations on an ADDG is shown in Fig. 3, where, operators, $\oplus$ is associative, $\otimes$ is commutative and $\circledast$ is both commutative and associative.

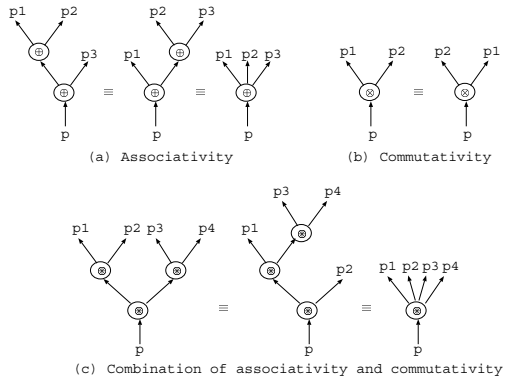

**Fig. 3: Effect of algebraic transformations on an** ADDG.

*Associativity.* As shown in Fig. 3(a), the end-nodes are regrouped with respect to the chain of $\oplus$-nodes (the *associative chain*) by the associative transformation, while maintaining their order.

*Commutativity.* As shown in Fig. 3(b), the effect of a commutative transformation is to permute the positions of the outgoing edges of the $\otimes$-node.

*Combination of associativity and commutativity.* As shown in Fig. 3(c), the effect of the transformation is to maintain the same end-nodes with any possible tree of $\circledast$-nodes between them and the root $\circledast$-node.

The designers, based on the knowledge of the overall computation, are often able to apply algorithmic or big-step transformations. For example, as can be seen from the ADDGs in Fig. 2, the transformed functions (c) and (d) are the result of several applications of the basic algebraic transformations above. Here, the transformation is motivated by the observation that they perform $N/2$ integer additions less when compared to functions (a) and (b) (i.e., $3N - 5N/2$).

## 5 Equivalence Checking Method

The method shows equivalence of the original and the transformed program functions by checking a sufficient condition. The condition is that, in the ADDGs extracted from the two program functions, every corresponding data dependence paths have identical ① *computation* and ② *output-input mappings*. The method checks these two parts of the sufficient condition based on a traversal of the two ADDGs.

We first briefly explain the basic method that is able to handle only expression propagations and loop transformations in Section 5.1. In Section 5.2, we extend it to handle algebraic transformations along with the other two transformations.

### 5.1 Basic method

Central to our equivalence checking is a *synchronized depth-first traversal* of the two ADDGs. The traversal begins from each of the corresponding root nodes (output variables with same names) and proceeds in lock-step on the two ADDGs. Initially the output-current mapping is set to identity mapping from an output variable to itself, for all its elements.

When an intermediate variable is encountered on the path in either of the two ADDGs, it is reduced and the output-current mappings are updated for each of the outgoing paths in the sub-ADDG rooted at the node being reduced. When an operator node is reached on one of the ADDGs, the same operator node must be reached next on the other ADDG, possibly after a sequence of intermediate variable reductions. Whenever multiple outgoing branches are present, the paths on the two ADDGs are paired for further traversal. If the branching is at an operator node, the paths with the same labels on the outgoing edges on either side are paired. If the branching is at an intermediate variable, the pairing is based on the output-current mapping. At any given point during the traversal, the paths already traversed on the two ADDGs are both guaranteed to have the same operator nodes appearing in the same sequence on them. This implies that, when a path ends at a leaf node, the same computation is guaranteed on the corresponding paths traversed in the two ADDGs. This satisfies the first part of the sufficient condition.

When the output-current mappings are updated at the leaf nodes on the two corresponding paths we have their output-input mappings. The second part of the condition is satisfied when they are checked to be identical. This implies that corresponding paths supply the same operators with the same values in any execution of the function. When the traversal has exhausted all the paths, satisfying the sufficient condition for each path, the two ADDGs, hence the two programs, are both guaranteed to apply the same computation on the same input values and hence assign the same output values.

For example, consider the ADDGs of functions (a) and (b) in Fig. 2, where (b) has been obtained by applying only expression propagations and loop transformations. There are 4 paths in total from the output variable C to the input variables in (a). But in (b), assignment to C is distributed among statements t3 and t4, as a result, it has 8 paths. For both (a) and (b), if we number the paths from left to right, the traversal corresponds, path 1 in (a) to paths 1 and 5 in (b), path 2 in (a) to paths 2 and 6 in (b) and so on. This satisfies first part of the sufficient condition since corresponding paths are found without any mismatch. It can also be checked that the updated output-input mappings on all pairs of corresponding paths are identical. For instance, for path 1 in (a) and (b), we have the following identical mappings.

[a]$M_{C \rightsquigarrow B}^1 \Leftrightarrow$ [b]$M_{C \rightsquigarrow B}^1 \Leftrightarrow \{[k] \to [2k] \mid 0 \leq k < 512 \wedge k \in \mathbb{Z}\}$

Note that since the branching at C in (b) divides the elements of C into two groups of assignments, output-input mapping is split for all paths of both (a) and (b). The remaining pairs of output-input mappings on the corresponding paths are similarly identical. This satisfies the second part of the sufficient condition.

### 5.2 Extended method

Let us now consider the situation when algebraic transformations are allowed. As discussed in Section 4, they may shuffle the paths and/or redistribute some operator nodes. Clearly, the traversal as described above will not suffice anymore. For example, consider the ADDGs (a) and (c) in Fig. 2, where (c) has been obtained by additionally applying algebraic transformations. In path 1 of the two ADDGs, a mismatch occurs upon reaching the input variable (leaf node) which is B in the original ADDG (a), whereas it is A in the transformed ADDG (c). The mismatch prevents the equivalence proof of program pairs under algebraic transformations.

In order to handle algebraic transformations, the traversal has to do additional work to know which paths to pair up. This requires that upon reaching an operator which permits algebraic transformations, a specific normal form be established before continuing the traversal. Such a normalization relies on two operations viz., *flattening* and *matching*, invoked depending on the properties that hold for the operator. When the operator is associative, flattening is invoked and when the operator is commutative, matching is invoked. When the operator is both associative and commutative, a flattening operation is followed by a matching operation.

**Flattening for an associative operator.** Suppose that an associative operator ($\oplus$) is reached on an ADDG. The flattening operation involves a lookahead traversal of the sub-ADDG of the associative chain rooted at the $\oplus$-node. It constructs an ordered list of nodes such that each node in the list is either a leaf node or the first node for an outgoing path which is an operator node different from $\oplus$. Any intermediate variables that exist on the path between the nodes in the list and the root $\oplus$-node are reduced. The effect of flattening is that it brings all operands of the chain to the same level as successor nodes of the root $\oplus$-node. Fig. 4 illustrates this. The order in which the operands are reached during the traversal is maintained by labeling the edges accordingly.

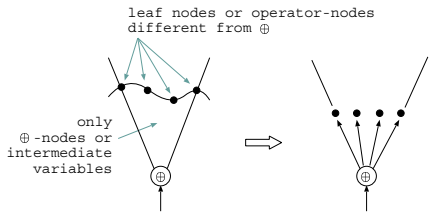

**Fig. 4: Illustration of the flattening operation.**

**Matching for a commutative operator.** Suppose that a commutative operator ($\otimes$) is reached on both the ADDGs. Since any permutation of the outgoing edges of $\otimes$-nodes on the two ADDGs is a valid transformation, the labels on them are of no consequence in pairing the corresponding paths. The correct pairing is then provided by the matching operation. Separately on both the ADDGs, the sub-ADDG rooted at the $\otimes$-node is traversed and any intermediate variables that are present are recursively reduced until all the successor nodes of the $\otimes$-node are either leaf nodes or operator nodes. This yields two lists of successor nodes, one for each ADDG. If the lists have unique nodes, then the pairing of the paths is one-to-one. Otherwise, if the non-unique nodes are input variables, output-input mappings are checked to pair them. If the non-unique nodes are operators, a lookahead traversal is recursively employed to reveal more successor nodes until a unique pairing is obtained.

Going back to our example function pair (a) and (c), we see that the first operator starting from the output variable is an addition operator. Therefore the flattening operation is applied at the node. This results in the ADDGs shown in Fig. 5. A subsequent matching operation here has to deal with the non-unique input variables as the successor nodes. Nodes B and A each appear twice as successor nodes on each of the ADDGs. This requires that the matching be established by checking the output-input mappings to the nodes.

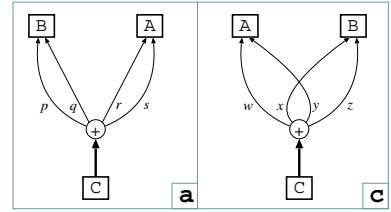

**Fig. 5: The ADDGs of functions (a) and (c) after flattening.**

The check reveals that the following equalities hold, where $D := \{[k] \mid 0 \leq k < 1024 \land k \in \mathbb{Z}\}$.

$$^a M_{C \stackrel{p}{\leadsto} B} \Leftrightarrow {^c M_{C \stackrel{z}{\leadsto} B}} \Leftrightarrow \{[k] \to [2k] \mid k \in D\}$$
$$^a M_{C \stackrel{q}{\leadsto} B} \Leftrightarrow {^c M_{C \stackrel{x}{\leadsto} B}} \Leftrightarrow \{[k] \to [k] \mid k \in D\}$$
$$^a M_{C \stackrel{r}{\leadsto} A} \Leftrightarrow {^c M_{C \stackrel{y}{\leadsto} A}} \Leftrightarrow \{[k] \to [2k] \mid k \in D\}$$
$$^a M_{C \stackrel{s}{\leadsto} A} \Leftrightarrow {^c M_{C \stackrel{w}{\leadsto} A}} \Leftrightarrow \{[k] \to [k] \mid k \in D\}$$

This results in the matching: $\{(p,z),(q,x),(r,y),(s,w)\}$. Here, a leaf node has been reached now on each of the paths and there exists an identical output-input mapping on the corresponding paths of the two ADDGs. Our sufficient condition therefore implies that they are equivalent.

To summarize, with the help of the flattening and matching operations, the synchronized traversal can be continued with a correct pairing of the branching paths at the operators which permit algebraic transformations. On each path, the traversal culminates at the matching leaf nodes, at which point, the second part of the sufficient condition is checked. That is, the output-input mapping computed on the path to that input variable during the traversal must be identical to the one on the corresponding path on the other ADDG. If this check fails, the traversal stops, reporting a failure and generating diagnostics. If it succeeds, the traversal continues until all the paths of the ADDGs are exhausted.

Before concluding, a brief remark is in order related to presence of any cycles in an ADDG. A cycle implies that the data-flow has *recurrences*, that is, a set of statements that read values (possibly computed from values –) written by themselves in earlier iterations. The method efficiently deals with cycles by employing the computation of *transitive closure* of the total dependence mapping of the cycle. This is computable only under certain conditions that usually hold in most real-life programs that we have checked on.

## 6 Verification and Debugging

Our prototype transformation verification tool implements the scheme shown in Fig. 6. The tool accepts original and transformed functions in the C language. Our sufficient condition assumes that the code is correctly scheduled. Therefore, it is required to check separately that all the reads for values follow their writes, that is, the def-use order is correct in the two programs. This can be checked by standard array data-flow analysis (cf. [1]). If the order is correct, the tool extracts ADDGs based on a source code analysis. The equivalence checker takes these ADDGs and applies the method we have presented. It relies on the OMEGA calculator [8] for an efficient implementation of all the required operations on integer sets and tuple relations. If the checking succeeds, the two program functions are guaranteed to be functionally equivalent. If it fails, it generates error diagnostics.

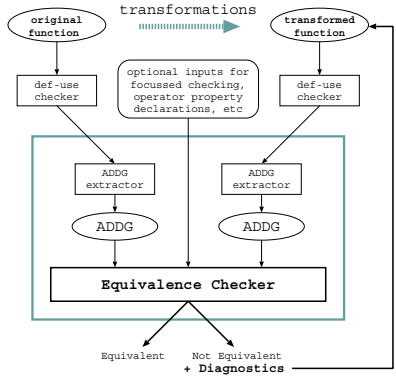

**Fig. 6: The verification and debugging scheme.**

### 6.1 Error diagnostics

The simplest case of an error is when there is a mismatch in the expected operator node or the leaf node. The information about the mismatch and the statement number is output, which suffices to correct the error. The more complicated errors involve the index expression. For example, we earlier mentioned that function (d) is in error. Let us see the diagnostics generated for it. It is similar to (c), but has the assignment to C interleaved among statements v3 and v4. Therefore flattening ADDG of function (d) results in the same ADDG as shown in Fig. 5(c), for each of the two branches at C. As explained earlier, the presence of non-unique leaf nodes requires that the matching, between the flattened ADDGs (a) and (d), be based on the dependency mappings. The matching succeeds for two paths, viz., $\{(q,x),(s,w)\}$, but fails for the other two paths, viz., $\{(p,z),(r,y)\}$. This is shown below.

$$(^{a}M_{C \overset{p}{\leadsto} B} \Leftrightarrow \{[x] \to [2x]\}) \quad \not\Leftrightarrow \quad (^{d}M_{C \overset{z}{\leadsto} B} \Leftrightarrow \{[x] \to [x]\})$$
$$(^{a}M_{C \overset{r}{\leadsto} A} \Leftrightarrow \{[x] \to [2x]\}) \quad \not\Leftrightarrow \quad (^{d}M_{C \overset{y}{\leadsto} A} \Leftrightarrow \{[x] \to [x]\})$$

where, $x \in \{[k] \,|\, (\exists j \,|\, 2j = k \land 0 \leq k < 1023 \land k \in \mathbb{Z})\}$. The above mismatch in dependency mappings implies that paths $z$ and $y$ are in error, which correspond to statements v3 and v1 in the program text. The diagnostic points the user to these two statements, displays the index expressions of variables C, $buf_2$, A and B in the statements as possible places of error and the difference in the output-input mappings. A further heuristic on this information deduces that variable $buf_2$ is common to the two paths and hence its index expression is most likely to be in error. This is indeed the case in statement v3 of function (d), where, it should have been buf[2*k].

When desired, the designer can also limit the checker to focus on only certain parts of the input programs. This can be done by specifying the subsets of output and input variables, or by declaring a correspondence of intermediate variables in the two programs. This helps not only in reducing the checking time but also in generating better error diagnostics.

### 6.2 Experience

The method is based on the depth-first traversal of the ADDGs and it uses tabling of established equivalences to avoid reworking on any overlapping sub-ADDGs. Therefore the complexity of traversal is linear in the size of the larger of the two ADDGs. Also, the supposedly expensive operations on the integer sets and tuples can be safely assumed to be bound by a small constant as the lengths of the formulae describing them are usually small enough in practice.

We have earlier reported verification times taken by our tool that implemented the basic method to be in the order of only few seconds [11]. In experiments with the new tool implementing the extended method, on realistic examples involving algebraic transformations, we have observed no significant degradation in performance. On problem instances that we experimented on, where we used source codes whose control complexity and ADDG sizes were comparable to real-life application kernels, verification consistently took less than 100 seconds on a desktop. This shows that our verification tool can be conveniently used to increase designer productivity while applying source code transformations.

## 7 Conclusions

We have presented a verification tool as required by designers applying source code transformations on signal processing programs. The tool is fully automatic, fast and able to provide useful error diagnostics. It is based on equivalence checking of the original and the transformed programs and is able to handle important transformations like expression propagations, loop and algebraic transformations that have been widely reported in the literature. Allowing algebraic transformations has significantly increased the class of programs that can be shown equivalent by the tool and hence its applicability in practice.

**Acknowledgment.** The authors gratefully thank anonymous referees for their detailed constructive comments.

## References

[1] R. Allen and K. Kennedy. *Optimizing Compilers for Modern Architectures*. Morgan Kaufmann Publishers, 2001.
[2] D. Barthou, P. Feautrier, and X. Redon. On the equivalence of two systems of affine recurrence equations. *8th Euro-Par*, LNCS 2400, pp. 309–313. Springer, 2002.
[3] C. Brandolese et al. The impact of source code transformations on software power and energy consumption. *Journal of Circuits, Systems, and Computers*, 11(5):477–502, 2002.
[4] F. Catthoor et al. *Custom Memory Management Methodology: Exploration of Memory Organization for Embedded Multimedia System Design*. Kluwer, 1998.
[5] F. Catthoor et al. *Data Access and Storage Management for Embedded Programmable Processors*. Kluwer, 2002.
[6] P. Feautrier. Array expansion. *International Conference on Supercomputing*, pp. 429–441. ACM, 1988.
[7] B. Goldberg et al. Into the loops: Practical issues in translation validation for optimizing compilers. *3rd Compiler Optimization Meets Compiler Verification*, ENTCS. Elsevier, 2004.
[8] W. Kelly et al. *The Omega Calculator and Library, Version 1.1.0*. http://www.cs.umd.edu/projects/omega.
[9] N. Mateev, V. Menon, and K. Pingali. Fractal symbolic analysis. *ACM TOPLAS*, 25(6):776–813, 2003.
[10] G. C. Necula. Translation validation for an optimizing compiler. *SIGPLAN PLDI*, pp. 83–95. ACM, 2000.
[11] K. C. Shashidhar et al. Automatic functional verification of memory oriented global source code transformations. *8th High Level Design Validation and Test*, pp 31–36. IEEE, 2003.
[12] R. A. van Engelen and K. A. Gallivan. An efficient algorithm for pointer-to-array access conversion for compiling and optimizing DSP applications. *Innovative Archs. for Future Gen. High-Perf. Processors and Systems*, pp. 80–89. IEEE, 2001.
[13] W. Wolf and M. Kandemir. Memory system optimization of embedded software. *Proc. of the IEEE*, 91(1):165–182, 2003.